# Non-intrinsic superconductivity in InN epilayers: role of Indium Oxide

Abdul Kadir[1,a], Sourin Mukhopadhyay[1], Tapas Ganguli[2], Charudatta Galande[1], M. R. Gokhale[1], B.M. Arora[1], Pratap Raychaudhuri[1] and Arnab Bhattacharya[1]

[1]Dept. of Condensed Matter Physics and Materials Science, Tata Institute of Fundamental Research, Homi Bhabha Road, Mumbai 400005, India
[2] Solid State Laser Division, Raja Ramanna Centre for Advanced Technology, Indore 425013, India

*Abstract:*
In recent years there have been reports of anomalous electrical resistivity and the presence of superconductivity in semiconducting InN layers. By a careful correlation of the temperature dependence of resistivity and magnetic susceptibility with structural information from high-resolution x-ray diffraction measurements we show that superconductivity is not intrinsic to InN and is seen only in samples that show traces of oxygen impurity. We hence believe that InN is not intrinsically a superconducting semiconductor.

a) email: abdulkadir@tifr.res.in

Epitaxial InN layers have been investigated extensively in recent years due to the controversy over its bandgap value, as well as the excellent electrical properties theoretically predicted for this material. For example, among the group III-Nitrides, InN has the highest electron mobility, and peak overshoot velocity, and the least dependence of bandgap on temperature[1]. Extensive experimental effort leading to an improvement in material quality has led to near-consensus on the value of the fundamental bandgap of good quality InN to be ~0.7 eV (e.g. Refs. 1,2). There is, however, still relatively lesser experimental data on the electrical properties of the material. There have been sporadic reports in the literature of the presence of superconductivity in epitaxially grown InN layers. Muira et al.[3] observed anomalous electrical properties in epitaxial InN films, where they observed a sharp decrease in the resistivity below 4.2K. Since neither indium droplets nor indium precipitation were observed by the AFM images and x-ray diffraction rocking curves, they attributed this drastic change in the resistivity as the indication of the occurrence of a phase transition to a possibly superconducting state. Many of the reports [4-6] on superconductivity in InN including measurements of superconducting behavior as a function of carrier density and discussions on possible mechanisms are from the group of Inushima et al. They claimed the superconductivity to be anisotropic and Type-II and suggest that neither the surface electron accumulation layer nor the metal-In precipitation has any contribution to the superconductivity. Inushima et al. postulate that superconductivity in InN is seen only in films within an optimum carrier density range. The lowest carrier density is limited by the Mott transition, at $n_e \sim 2 \times 10^{17}$ cm$^{-3}$ and the highest density is limited by the superconductivity to metal transition at $n_e \sim 5 \times 10^{20}$ cm$^{-3}$. Within this range there are nano-sized In-In chains along the [11-20] direction resulting from inversion domains of InN grown on sapphire, which form micro-scale Josephson junctions[7,8]. In another study Chang et al.[9] also reported superconductivity in epitaxially grown InN films where however, they rule out In-enriched networks contributing to the superconducting behavior from data based on Raman shifts. Most of these papers have focused on the study of superconducting behavior and have not made a detailed study of the link with the structural properties of the layer. The cause of superconductivity in InN has been difficult to establish, and there is still no conclusive evidence of the mechanisms that have been proposed.

We have earlier reported[10] a detailed study of the growth parameter space for InN in a close-coupled showerhead metalorganic vapor phase epitaxy (MOVPE) system, examining the effects of V/ III ratio, temperature, reactor pressure, precursor flux, etc. As a result of this study covering over 40 growth experiments, we have access to a range of InN samples of varying quality[11]. Some of these samples show superconducting behavior at different transition temperatures, in other samples there is no evidence of superconductivity till the lowest temperatures measured. In this letter we correlate these observations with structural data from high-resolution x-ray diffraction and experiments on annealed InN layers to show that the presence of residual oxygen impurities is responsible for the behavior seen.

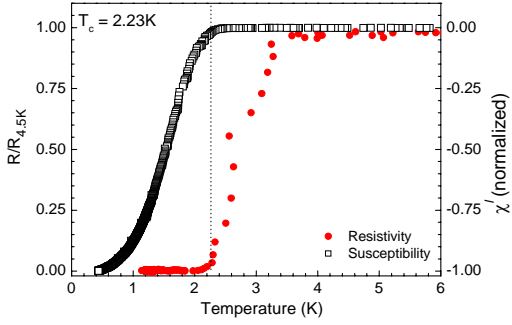

*Fig. 1: Normalized resistance and a.c. susceptibility for the InN sample C showing superconducting transition at 2.3K. As expected, the onset of the transition in susceptibility coincides with the temperature at which the resistance falls to 5% of its normal state value.*

All the InN epilayers studied in this work were deposited by MOVPE on 2" *c*-plane sapphire substrates in a 3x2" close-coupled-showerhead reactor (Thomas Swan) using trimethylindium (TMIn) and ammonia ($NH_3$) as precursors with nitrogen as the carrier gas. The InN layers were 0.2-0.4 μm thick and grown on a 1-μm-thick undoped GaN buffer layer. The details of the growth have been discussed previously[10]. The InN films were structurally characterized by high-resolution X-ray diffraction (HRXRD) on a PANalytical X-pert MRD system with a Hybrid 4-bounce monochromator at the input having a divergence of ~20 arc seconds. The superconductivity data were obtained from dc transport and ac susceptibility measurements carried out in custom-built $He^4$ and $He^3$ cryostats in the temperature range 0.3K–300 K. The resistance was measured in the standard 4-probe configuration. The high frequency (15kHz) ac susceptibility setup consisted of two planar coils (acting as the primary and secondary, respectively) between which the sample was sandwiched. When the sample becomes superconducting, it shields the magnetic field produced by the primary from the secondary, and causes a sharp drop in the real part of the signal picked up by the secondary coil.

Fig. 1 shows typical superconducting transition characteristics for an InN epilayer. The vertical axis shows the normalized values of the resistivity and susceptibility and the data is plotted only till 15K for the sake of clarity. The residual resistance ratio ($R_{4.2K}/R_{300K}$) of all the films studies is <2 indicating that they behave like dirty metals. We determine the $T_c$ using two criteria: (i) the temperature where the resistance is 5% of its normal state value and (ii) the temperature where the ac susceptibility deviates from zero.

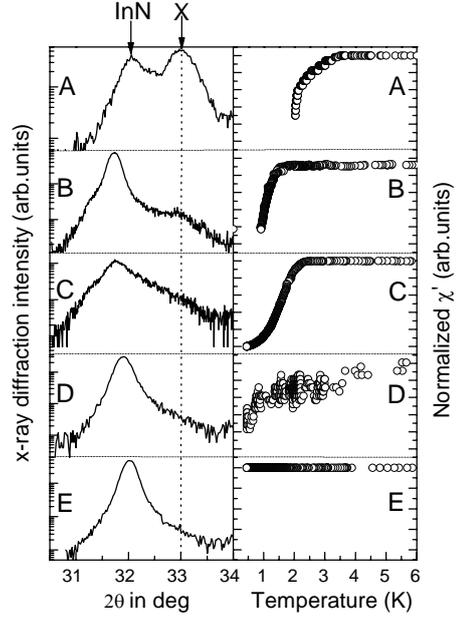

*Fig. 2: (left column) X-ray diffraction profile around the InN peak for the (0002) reflection in the ω/2θ scattering geometry showing the shoulder X to the right of the InN peak, and (right column) normalized real part of susceptibility (χ') as a function of temperature for a series of 5 representative InN samples (A-E).*

The InN samples studied show a variation in their superconducting behavior, with different transition temperatures seen for samples deposited under various growth conditions, with some samples not showing any superconducting behavior till the low temperature limit of the cryostat. The right column of Fig. 2 shows the measured susceptibility as a function of temperature from a series of 5 representative InN samples (A-E) where the superconducting transition temperature decreases from about 3.1K for sample A to 0.5K for D, and sample E does not show any superconductor transition. The left column of Fig 2 shows the region around the InN peak for the (0002) reflection in the ω/2θ scattering geometry from the same samples. A close observation reveals that the InN peaks in samples other than E are not symmetric on both sides, and on the side of higher 2θ value, a shoulder "X" is observed whose relative intensity changes from sample to sample. This feature increases gradually from sample D through A, concomitant with an increase in the measured superconducting transition temperature. Sample A, with the highest transition temperature shows a clear peak corresponding to the feature X. The data in Fig. 2 thus illustrates a close correlation between the relative intensity of the feature X and the superconducting transition temperature.

If InN is indeed a superconductor, it is surprising that the only select samples show a superconducting transition and that too at different transition temperatures. We hence believe that the superconductivity is related to incorporation of an impurity phase, which shows up as the feature X in the x-ray diffraction profile. It is interesting to note that the 2θ value corresponding to the feature X is around 33°. This strong "anomalous 33° feature" in the x-ray diffraction profile of InN has been a source of discussion[12,13] in the recent literature as there are many possible sources that may lead to this feature: free In[14,15], polycrystalline InN[16], further, cubic phases of InN, and cubic/rhombohedral phases of $In_2O_3$ also have Bragg reflections around 2θ = 33° [Ref. 17]. Of these, most groups have considered free Indium, with a $T_c$ of 3.4K to be a likely suspect candidate for super-conductivity. We have tried to ascertain if the peak seen at 2θ = 33°, which is near the In(10-11) reflection at 2θ = 32.964°, arises from free In. This peak is nearly oriented, as ascertained by reciprocal space maps around the InN peaks. Further free In also has additional peaks with moderate intensity at 2θ = 36.304° and 2θ = 39.155°, which should be easily detected in our x-ray diffractometer. However we are unable to see any other reflections corresponding to free metallic Indium. There is also no change in the x-ray diffraction pattern on annealing the sample well beyond the melting point of Indium. From this, we conclude that free indium is not responsible for the feature X. Further, reciprocal lattice maps of InN (0002) and InN (10-11) reflections show that the structure corresponding to feature X does not have the same symmetry as InN, which also rules out InN being responsible for feature X.

We have performed a set of annealing experiments on these InN films to help understand the nature of the feature X. Annealing experiments were performed in a nitrogen environment in a Rapid Thermal Annealing system, which permitted the introduction of small quantities of air. The InN samples were annealed at the growth temperature (530°C) for 4 minutes and then at 650°C, just near the dissociation temperature of InN[18], for 2.5 minutes, and the x-ray diffraction profile measured thereafter. Fig 3a shows the comparison of x-ray diffraction measurement from the sample B before and after annealing at 650°C. As a result of annealing the x-ray peak intensity from InN is reduced due to the dissociation of InN. Furthermore, the anonymous feature, which appeared as a shoulder before annealing, has now become a distinct peak. Together with this, one more peak has appeared at a 2θ value of 30.62°, which does not match with any of the metallic indium peaks. The JCPDS data shows that these peak positions actually match closely the 2θ values of $In_2O_3$ of the (110) reflection of a rhombohedral phase (2θ = 30.62°) and (222) reflection of a cubic phase (2θ = 32.92°). Thus, we conjecture that the feature X arises from the above-mentioned two phases of $In_2O_3$. This provides a possible explanation for the superconducting nature of the samples. Indium oxide is a well-known superconductor both in granular as well as in amorphous form[19,20]. The superconducting transition temperature of indium oxide strongly depends on the nature of the sample. Granular $In_2O_3$ has been reported to show a $T_c$~3.2K which is close to the $T_c$ observed in sample A. Amorphous indium oxide on the other hand has been reported to exhibit a broad range of $T_c$ varying from 1.9K down to 300mK depending on the preparation and annealing conditions[19,20]. To confirm this conjecture, the annealing experiment was repeated on the non-superconducting sample E. Fig 3b compares the x-ray diffraction profiles of the sample E before and after the annealing at 650°C. On annealing, we see two new peaks appearing at exactly the same 2θ values corresponding to $In_2O_3$, and matching the shoulder of feature X. Further, the resistance as a function of temperature for sample E, before and after annealing is shown in Fig. 3c. After annealing, the sample shows a broad superconducting transition, with a $T_c$ ~3.1K. This clearly shows that the superconductivity in InN films arises from the presence of $In_2O_3$ impurities formed during the process of annealing. We have attempted to measure the amount of oxygen present in the InN samples via secondary ion mass spectrometry. While an absolute quantification is difficult without the availability of standard oxygen implanted reference samples for InN, we can clearly see a very high level of oxygen in superconducting samples compared to the non-superconducting ones.

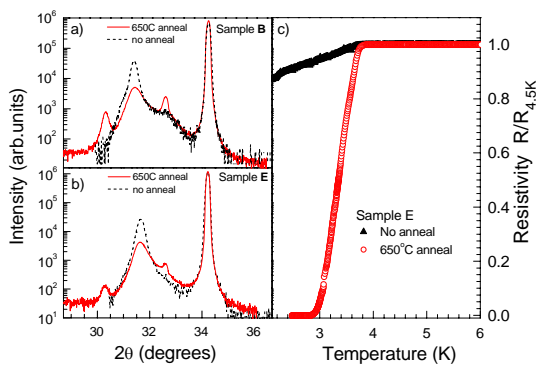

*Fig. 3: (a) and (b) X-ray diffraction profiles of samples B and E and (c) Resistance vs. temperature for sample E, before and after annealing at 650°C. See text for details.*

In summary, we believe the above set of experiments on the series of InN samples clearly establishes that the superconductivity seen is not intrinsic to the InN, and arises from the presence of trace amounts of indium oxide in the sample. Depending on the amount of amount and the nature of indium oxide present, the transition temperature varies over from 3.1K to 0.5K. However, no superconductivity is found in samples where we do not observe any trace of indium oxide.